\documentclass[11pt,a4paper]{article}

\usepackage[utf8]{inputenc}
\usepackage[T1]{fontenc}
\usepackage[english]{babel}

\usepackage[
  top=1in,
  bottom=1in,
  left=1in,
  right=1in
]{geometry}

\usepackage{lmodern}
\usepackage{microtype}

\usepackage{amsmath,amssymb,amsfonts,mathtools}
\usepackage{bm}

\usepackage{booktabs}
\usepackage{array}
\usepackage{multirow}
\usepackage{graphicx}
\usepackage{float}
\usepackage{caption}
\usepackage{tikz}
\usetikzlibrary{positioning,arrows.meta,calc}

\usepackage{enumitem}
\usepackage{xcolor}
\usepackage{xspace}

\usepackage[numbers,sort&compress]{natbib}

\usepackage[
  colorlinks=true,
  linkcolor=blue!70!black,
  citecolor=green!50!black,
  urlcolor=blue!60!black
]{hyperref}
\usepackage{cleveref}

\DeclareMathOperator{\sign}{sign}

\newcommand{\R}{\mathbb{R}}

\title{%
  Multifractal Flexibly Detrended Fluctuation Analysis\\
  as Feature Engineering for Attention-Augmented LSTM\\
  Commodity Price Forecasting%
}

\author{
  Ahmad Nasser\textsuperscript{1}\quad
  Syed M.\ Abdullah\textsuperscript{2,}\thanks{Corresponding author: \texttt{s.muhammad.abdullah67@gmail.com}}%
  \\[6pt]
  \textsuperscript{1}Society of Economics, St.\ Christopher School of Bahrain\\
  \textsuperscript{2}Department of Economics, Lahore University of Management Sciences (LUMS)
}

\date{October 20, 2026}

\begin{document}
\maketitle

\begin{abstract}
This paper investigates whether rolling multifractal regime descriptors,
computed via Multifractal Flexibly Detrended Fluctuation Analysis
(MFFDFA), can improve deep-learning-based commodity return forecasting.
We implement a vectorised MFFDFA algorithm following Rak and
Zi\k{e}ba~(2015) and extract rolling estimates of the generalised Hurst
exponent~$h(q)$, spectrum width~$\Delta\alpha$, and spectrum asymmetry
as auxiliary input features for an attention-augmented Long Short-Term
Memory (LSTM) network. The hybrid model is compared against a baseline
LSTM trained on standard price-based features, using identical
architecture, hyperparameters, and chronological train/validation/test
splits across multiple commodity series. A multi-seed ensemble protocol
and the Diebold--Mariano test are employed to assess statistical
significance. We additionally compare MFFDFA with classical fixed-order
polynomial MFDFA and apply surrogate shuffle tests to verify that the
detected multifractality reflects genuine temporal correlations rather
than distributional artefacts.

\end{abstract}

\section{Introduction}\label{sec:introduction}

Commodity prices play a central role in global economic systems,
influencing inflation, trade balances, and monetary policy decisions
across both developed and emerging
economies~\cite{byrne2013commodity}. Accurate forecasting of commodity
price movements is therefore of substantial interest to policymakers,
traders, and risk managers alike. However, commodity price series are
notoriously difficult to predict, exhibiting complex nonlinear dynamics,
heavy-tailed distributions, long-range dependence, and time-varying
volatility that severely violate the assumptions of classical linear
models~\cite{mandelbrot1963variation,cont2001empirical}.

A growing body of evidence suggests that commodity price series exhibit
\emph{multifractal} behaviour---that is, their statistical scaling
properties are not uniform but vary across different time scales and
fluctuation magnitudes~\cite{kantelhardt2002multifractal,zunino2008multifractal}.
Unlike monofractal processes, which are fully characterised by a single
Hurst exponent, multifractal series require a spectrum of local scaling
exponents to describe their complexity. The presence of multifractality
in financial time series has been documented extensively for equity
markets~\cite{jiang2019multifractal}, foreign exchange
rates~\cite{norouzzadeh2006multifractal}, and commodity markets
including crude oil~\cite{he2011multifractal},
gold~\cite{mali2014multifractal}, and agricultural
products~\cite{wang2014market}.

\textbf{Multifractal Detrended Fluctuation Analysis} (MFDFA), introduced
by Kantelhardt et~al.~\cite{kantelhardt2002multifractal} as a
generalisation of the Detrended Fluctuation Analysis (DFA) method of
Peng et~al.~\cite{peng1994mosaic}, has become the standard tool for
characterising the multifractal properties of nonstationary time series.
In MFDFA, the series is divided into non-overlapping segments, each
detrended by a fixed-order polynomial, and the resulting fluctuation
function is computed across a range of moment orders~$q$ and segment
sizes~$s$. The log-log scaling relationship between the fluctuation
function and segment size yields the generalised Hurst exponent~$h(q)$,
from which the singularity spectrum~$f(\alpha)$ is derived via the
Legendre transform.

However, the use of a fixed polynomial order for detrending is a
well-recognised limitation of classical
MFDFA~\cite{oswiecimka2006wavelet}. The local trend structure of a
financial time series may vary across segments---some segments may be
well-approximated by a quadratic trend, while others may exhibit more
complex local dynamics. To address this limitation, Rak and
Zi\k{e}ba~\cite{rak2015mffdfa} proposed \textbf{Multifractal Flexibly
Detrended Fluctuation Analysis} (MFFDFA), in which a set~$\mathcal{Q}$
of candidate detrending functions is applied to each segment, and the
function yielding the highest coefficient of determination~($R^2$) is
retained. This adaptive, per-segment detrending yields more accurate
estimates of the multifractal spectrum, particularly in heterogeneous
real-world signals. Additionally, the MFFDFA framework incorporates
overlapping segments controlled by an overlap parameter~$k$, improving
the statistical stability of fluctuation estimates at each scale.

On the forecasting side, \textbf{Long Short-Term Memory} (LSTM)
networks~\cite{hochreiter1997long} have emerged as powerful tools for
sequential prediction tasks, including financial time series forecasting.
Unlike traditional recurrent neural networks, LSTMs overcome the
vanishing gradient problem through gated memory cells, enabling them to
capture long-range temporal dependencies. Recent work has further
enhanced LSTM architectures with \textbf{attention
mechanisms}~\cite{bahdanau2015neural}, which allow the model to
dynamically weight the importance of different time steps in the input
sequence rather than relying solely on the final hidden state.
Attention-augmented LSTMs have demonstrated superior performance in
financial forecasting by effectively filtering noise and focusing on the
most informative historical
observations~\cite{qin2017dual,li2019ealstm}.

Despite the extensive literature on both multifractal characterisation
and deep-learning-based forecasting separately, \emph{few studies have
explicitly investigated whether multifractal regime descriptors can
serve as informative features for deep learning models} in the context
of commodity price prediction. The central hypothesis of this study is
that providing an LSTM with rolling estimates of the generalised Hurst
exponent~$h(q)$, the spectrum width~$\Delta\alpha$, and spectrum
asymmetry---quantities that describe the current persistence regime and
scaling complexity of the series---can improve forecasting performance
beyond what is achievable with standard price-based features alone.

\medskip
\noindent This paper makes the following contributions:
\begin{enumerate}[label=(\roman*),leftmargin=*]
  \item A \textbf{vectorised implementation of MFFDFA} with overlapping
    segments, batch ordinary least squares, and a surrogate shuffle test
    for multifractality significance, following Rak and
    Zi\k{e}ba~\cite{rak2015mffdfa}.
  \item A \textbf{systematic comparison of classical MFDFA versus
    MFFDFA} on commodity price series, reproducing and extending the
    methodological findings of the original paper.
  \item A \textbf{hybrid forecasting framework} that uses rolling MFFDFA
    features as auxiliary inputs to an attention-augmented LSTM, compared
    against a baseline LSTM trained on standard price and return
    features.
  \item A \textbf{rigorous evaluation protocol} employing multi-seed
    ensemble averaging, chronological train/validation/test splits, and
    formal statistical testing to assess the significance of any
    performance differences.
\end{enumerate}

The remainder of the paper is organised as follows.
\Cref{sec:methodology} describes the methodology, including the MFFDFA
algorithm, the LSTM architecture, and the experimental design.
\Cref{sec:data} describes the PMEX data. \Cref{sec:results} presents the empirical results across multiple
commodity series. \Cref{sec:discussion} discusses the findings and their
implications, and \Cref{sec:conclusion} concludes.

\section{Methodology}\label{sec:methodology}

This section presents the analytical framework:
(\ref{subsec:mffdfa})~the Multifractal Flexibly Detrended Fluctuation
Analysis (MFFDFA) method used to characterise the scaling properties of
commodity return series;
(\ref{subsec:rolling})~the rolling extraction of MFFDFA features;
(\ref{subsec:baseline_features})~the baseline features;
(\ref{subsec:lstm})~the attention-augmented Long Short-Term Memory
(LSTM) architecture employed for one-step-ahead return prediction;
(\ref{subsec:training})~the training procedure;
(\ref{subsec:metrics})~the evaluation metrics; and
(\ref{subsec:experimental})~the experimental design for comparing
baseline and hybrid models.

\subsection{Multifractal Flexibly Detrended Fluctuation Analysis (MFFDFA)}
\label{subsec:mffdfa}

The MFFDFA method, proposed by Rak and
Zi\k{e}ba~\cite{rak2015mffdfa}, extends the classical MFDFA framework
of Kantelhardt et~al.~\cite{kantelhardt2002multifractal} by replacing
the fixed-order polynomial detrending with an adaptive, per-segment
function selection. The algorithm proceeds as follows.

\paragraph{Step~1 --- Profile construction.}
Given a time series of log returns~$\{x_i\}_{i=1}^{N}$, the cumulative
profile is computed as
\begin{equation}\label{eq:profile}
  Y(j) = \sum_{i=1}^{j}\bigl[x_i - \bar{x}\bigr],
  \quad j = 1, 2, \ldots, N,
\end{equation}
where $\bar{x}$ denotes the sample mean. This transformation converts
the stationary return series into a random-walk-like process suitable
for fluctuation analysis~\cite{peng1994mosaic}.

\paragraph{Step~2 --- Overlapping segmentation.}
The profile~$Y(j)$ is divided into segments of length~$s$ with a step
size of~$\lfloor s/k \rfloor$, where the overlap parameter
$k \geq 1$ controls the degree of overlap. Setting $k = 1$ recovers
the non-overlapping segmentation of classical MFDFA; higher values
of~$k$ increase the number of segments per scale, improving the
statistical stability of the fluctuation estimates. Following Rak and
Zi\k{e}ba~\cite{rak2015mffdfa}, we use $k = 2$ throughout this study.

\paragraph{Step~3 --- Flexible local detrending.}
For each segment~$\nu$ at scale~$s$, a set~$\mathcal{Q}$ of candidate
detrending functions is fitted by ordinary least squares (OLS). In this
study, $\mathcal{Q}$ consists of three functions, each linear in its
parameters despite being nonlinear in the independent variable~$t$:
\begin{table}[H]
  \centering
  \caption{Candidate detrending functions in the set~$\mathcal{Q}$.}
  \label{tab:candidates}
  \begin{tabular}{@{}llc@{}}
    \toprule
    \textbf{Name} & \textbf{Functional form} & \textbf{Design-matrix columns} \\
    \midrule
    Quadratic         & $f_1(t) = at^2 + bt + c$          & $[t^2,\; t,\; 1]$ \\
    Sinusoidal-linear & $f_2(t) = a\sin(t^2) + bt + c$    & $[\sin(t^2),\; t,\; 1]$ \\
    Cubic             & $f_3(t) = at^3 + bt + c$           & $[t^3,\; t,\; 1]$ \\
    \bottomrule
  \end{tabular}
\end{table}
\noindent For each candidate $f \in \mathcal{Q}$, the coefficient of
determination~$R^2$ is computed. The residuals from the function with
the highest~$R^2$ are retained for that segment. This per-segment
selection constitutes the ``flexible'' detrending that distinguishes
MFFDFA from classical MFDFA.

To achieve computational efficiency, the OLS fits across all segments at
a given scale are performed simultaneously using batch matrix operations.
Specifically, for a shared design matrix
$\mathbf{X} \in \R^{s \times p}$ and a data matrix
$\mathbf{Y} \in \R^{n_{\mathrm{seg}} \times s}$ (each row being one
segment), the coefficients are obtained by solving the normal equations
\begin{equation}\label{eq:normal}
  \bigl(\mathbf{X}^{\!\top} \mathbf{X}\bigr) \,\bm{\beta}
  = \mathbf{X}^{\!\top} \mathbf{Y}^{\!\top},
\end{equation}
eliminating the need for a loop over segments.

\paragraph{Step~4 --- Local variance.}
The local variance for segment~$\nu$ at scale~$s$ is
\begin{equation}\label{eq:local_var}
  F^2(\nu, s)
  = \frac{1}{s}\sum_{j=1}^{s}
    \Bigl[
      Y\!\bigl((\nu{-}1)\cdot\mathrm{step} + j\bigr)
      - f_{\nu}^{*}(j)
    \Bigr]^2,
\end{equation}
where $f_{\nu}^{*}$ denotes the best-fit function selected in Step~3
and $\mathrm{step} = \lfloor s/k \rfloor$.

\paragraph{Step~5 --- Fluctuation function and generalised Hurst
exponents.}
The $q$-th order fluctuation function is defined as
\begin{equation}\label{eq:fq}
  F_q(s) = \left\{
    \frac{1}{N_s}\sum_{\nu=1}^{N_s}
    \bigl[F^2(\nu,s)\bigr]^{q/2}
  \right\}^{1/q}, \quad q \neq 0,
\end{equation}
where $N_s$ is the number of segments at scale~$s$. For $q = 0$, the
logarithmic averaging limit is used:
\begin{equation}\label{eq:fq0}
  F_0(s) = \exp\!\left\{
    \frac{1}{2N_s}\sum_{\nu=1}^{N_s}
    \ln F^2(\nu,s)
  \right\},
\end{equation}
following the convention of Kantelhardt
et~al.~\cite{kantelhardt2002multifractal}. The generalised Hurst
exponent~$h(q)$ is obtained from the scaling relation
$F_q(s) \sim s^{h(q)}$ by linear regression of $\log F_q(s)$ against
$\log s$. The quality of each scaling fit is assessed via~$R^2$.

\paragraph{Step~6 --- Singularity spectrum.}
The multifractal spectrum is obtained via the Legendre transform:
\begin{align}
  \tau(q)    &= q \cdot h(q) - 1, \label{eq:tau} \\[4pt]
  \alpha     &= h(q) + q \cdot h'(q), \label{eq:alpha} \\[4pt]
  f(\alpha)  &= q\bigl[\alpha - h(q)\bigr] + 1, \label{eq:falpha}
\end{align}
where $h'(q) = \mathrm{d}h/\mathrm{d}q$ is estimated numerically. To
reduce noise in the Legendre transform, Savitzky--Golay smoothing
(window length~7, polynomial order~3) is optionally applied to~$h(q)$
prior to differentiation. The key summary statistics extracted from the
spectrum are:
\begin{itemize}[nosep]
  \item the Hurst exponent at $q = 2$, denoted~$h(2)$;
  \item the spectrum width
    $\Delta\alpha = \alpha_{\max} - \alpha_{\min}$, which quantifies
    the degree of multifractality;
  \item the spectrum asymmetry
    $A = (W_L - W_R)/(W_L + W_R)$, where $W_L$ and $W_R$ are the left
    and right half-widths of the spectrum relative to its peak.
\end{itemize}

\paragraph{Comparison with classical MFDFA.}
To evaluate the methodological contribution of the flexible detrending,
we additionally implement classical MFDFA with a single fixed-order
polynomial ($m = 2$) as the only detrending function, following the
standard formulation of Kantelhardt
et~al.~\cite{kantelhardt2002multifractal}. Both methods are applied to
the same series, and the resulting spectra are compared.

\paragraph{Surrogate shuffle test.}
To determine whether the observed multifractality arises from genuine
temporal correlations rather than from the marginal distribution of the
series (e.g., heavy tails), we perform a surrogate test. The series is
randomly shuffled to destroy all temporal structure while preserving the
marginal distribution, and MFFDFA is reapplied. This procedure is
repeated for $n = 19$ surrogates (yielding a minimum achievable
$p$-value of~0.05). If the spectrum width~$\Delta\alpha$ of the
shuffled series is comparable to that of the original, the
multifractality is attributable to distributional properties; if
$\Delta\alpha$ decreases significantly, the multifractality reflects
genuine temporal
correlations~\cite{kantelhardt2002multifractal}.

\subsection{Rolling MFFDFA Feature Extraction}\label{subsec:rolling}

To serve as inputs to the forecasting model, MFFDFA features are
computed on a rolling basis. At each time step~$t$, the MFFDFA
algorithm is applied to the trailing window of $W = 500$ trading days
(approximately two calendar years) of log returns. The features
extracted are summarised in~\Cref{tab:rolling_features}.

\begin{table}[t]
  \centering
  \caption{Rolling MFFDFA features extracted at each time step.}
  \label{tab:rolling_features}
  \begin{tabular}{@{}lcl@{}}
    \toprule
    \textbf{Feature} & \textbf{Symbol} & \textbf{Interpretation} \\
    \midrule
    Gen.\ Hurst exponent  & $h(2)$         & Persistence ($>0.5$) vs.\ anti-persistence ($<0.5$) \\
    Spectrum width         & $\Delta\alpha$ & Degree of multifractality / scaling heterogeneity \\
    Spectrum asymmetry     & $A$            & Dominance of small vs.\ large fluctuations \\
    Hurst at $q = -2$      & $h(-2)$        & Scaling of small fluctuations \\
    Hurst at $q = 0.5$     & $h(0.5)$       & Intermediate scaling \\
    Hurst at $q = 4$       & $h(4)$         & Scaling of large fluctuations \\
    \bottomrule
  \end{tabular}
\end{table}

To manage computational cost, features are recomputed every
$\Delta t = 10$ trading days and forward-filled between recomputation
points, reflecting the fact that the multifractal spectrum evolves
slowly relative to daily price movements.

\subsection{Baseline Features}\label{subsec:baseline_features}

Both the baseline and hybrid models receive the following standard
features computed from the price series~$\{P_t\}$:
\begin{itemize}[nosep]
  \item \textbf{Log return:}
    $r_t = \ln(P_t / P_{t-1})$;
  \item \textbf{Lagged returns:}
    $r_{t-1}$ and $r_{t-2}$ (1-day and 2-day lags);
  \item \textbf{Rolling volatility:}
    20-day rolling standard deviation of returns;
  \item \textbf{Momentum:}
    $\ln(P_t) - \ln(P_{t-10})$ (10-day log-price momentum).
\end{itemize}
The forecast target is the next-day log return~$r_{t+1}$.

\subsection{LSTM Architecture}\label{subsec:lstm}

\begin{figure}[t]
  \centering
  \begin{tikzpicture}[
      font=\footnotesize,
      >={Stealth[length=2mm]},
      blk/.style={draw, minimum width=6.6cm, minimum height=8mm,
                  align=center, inner sep=3pt},
      feat/.style={draw, minimum width=5.5cm, minimum height=11mm,
                   align=center, inner sep=3pt, font=\scriptsize},
      shp/.style={pos=0.5, right=1mm, font=\scriptsize}
    ]
    \node[feat] (base) at (-3.0,0)
      {\textbf{Baseline features} (5)\\
       $r_t,\ r_{t-1},\ r_{t-2}$, 20-day volatility,\\
       10-day momentum};
    \node[feat, dashed] (mf) at (3.0,0)
      {\textbf{Rolling MFFDFA features} (6)\\
       $h(2),\ \Delta\alpha,\ A,\ h(-2),\ h(0.5),\ h(4)$\\
       (hybrid model only)};
    \coordinate (mid) at ($(base)!0.5!(mf)$);
    \node[blk, below=9mm of mid] (inp)
      {Concatenate: input sequence of $L = 30$ trading days\\
       $d = 5$ (baseline) or $d = 11$ (hybrid)};
    \node[blk, below=6mm of inp] (proj)
      {Linear projection, $d \to H = 64$};
    \node[blk, below=6mm of proj] (lstm)
      {2-layer LSTM ($H = 64$, inter-layer dropout $0.2$)};
    \node[draw, circle, inner sep=1pt, below=6mm of lstm] (add) {$+$};
    \node[blk, below=6mm of add] (ln) {Layer normalisation};
    \node[blk, below=6mm of ln] (att)
      {Attention pooling\\
       $a_t = \mathrm{softmax}_t(\mathbf{w}^{\top}\mathbf{h}_t)$,\quad
       $\tilde{\mathbf{h}} = \sum_{t=1}^{L} a_t\,\mathbf{h}_t$};
    \node[blk, below=6mm of att] (head)
      {Prediction head\\
       Linear($H\!\to\!32$) $\to$ GELU $\to$ Dropout($0.2$)
       $\to$ Linear($32\!\to\!1$)};
    \node[below=6mm of head] (out) {$\hat{r}_{t+1}$ (next-day log return)};
    \draw[->] (base.south) -- (base.south |- inp.north);
    \draw[->] (mf.south)   -- (mf.south |- inp.north);
    \draw[->] (inp)  -- node[shp] {$L \times d$} (proj);
    \draw[->] (proj) -- node[shp] {$L \times H$} (lstm);
    \draw[->] (lstm) -- node[shp] {$L \times H$} (add);
    \draw[->] (add)  -- (ln);
    \draw[->] (ln)   -- node[shp] {$L \times H$} (att);
    \draw[->] (att)  -- node[shp] {$H$} (head);
    \draw[->] (head) -- node[shp] {$1$} (out);
    \coordinate (split) at ($(proj.south)!0.5!(lstm.north)$);
    \draw[->] (split) -- ++(-3.9,0) coordinate (rtop) |- (add.west);
    \node[left, font=\scriptsize] at ($(rtop)!0.5!(rtop |- add.west)$) {residual};
  \end{tikzpicture}
  \caption{Schematic of the attention-augmented LSTM forecaster.
    Input features are projected to dimension~$H$, processed by a
    two-layer LSTM with residual connections and layer normalisation,
    pooled via learnable attention, and mapped to a scalar predicted
    return. The dashed branch is present in the hybrid model only.}
  \label{fig:architecture}
\end{figure}
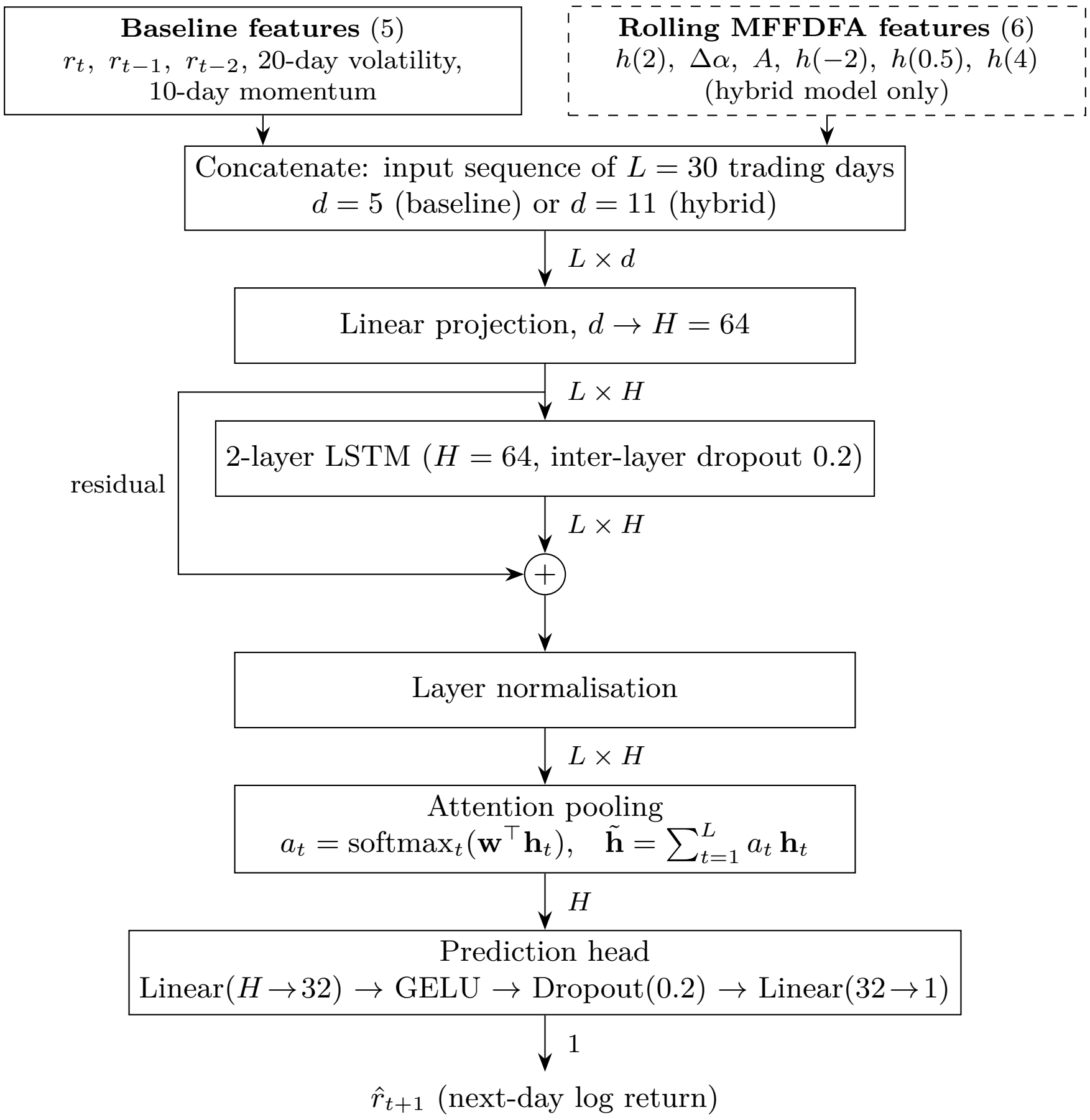

We employ an attention-augmented LSTM architecture for one-step-ahead
return prediction. The model comprises the components illustrated
schematically in~\Cref{fig:architecture} and detailed below.

\paragraph{Input projection.}
A linear transformation maps the raw feature vector of
dimension~$d$ to the hidden dimension $H = 64$, enabling residual
connections downstream.

\paragraph{LSTM encoder.}
A 2-layer LSTM with hidden dimension $H = 64$ and inter-layer dropout
of~0.2. The input to the LSTM is the projected feature sequence of
length $L = 30$ trading days (approximately six calendar weeks).

\paragraph{Residual connection and layer normalisation.}
The LSTM output is combined with the projected input via an additive
residual connection, followed by layer
normalisation~\cite{ba2016layer}. This stabilises training and
mitigates the degradation problem observed in deep sequential models.

\paragraph{Attention pooling.}
Rather than using only the final hidden state, a learnable attention
mechanism computes a weighted average across all time steps of the LSTM
output:
\begin{equation}\label{eq:attention}
  a_t
  = \frac{\exp\!\bigl(\mathbf{w}^{\!\top}\mathbf{h}_t\bigr)}%
         {\sum_{t'=1}^{L}\exp\!\bigl(\mathbf{w}^{\!\top}
           \mathbf{h}_{t'}\bigr)},
  \qquad
  \tilde{\mathbf{h}}
  = \sum_{t=1}^{L} a_t\,\mathbf{h}_t,
\end{equation}
where $\mathbf{w}$ is a learnable parameter vector and
$\mathbf{h}_t \in \R^{H}$ is the LSTM output at time step~$t$. This
allows the model to selectively attend to the most informative time
steps in the lookback window~\cite{bahdanau2015neural}.

\paragraph{Prediction head.}
The pooled representation~$\tilde{\mathbf{h}}$ is passed through a
two-layer feed-forward network:
\[
  \text{Linear}(H \!\to\! 32)
  \;\to\; \text{GELU}
  \;\to\; \text{Dropout}(0.2)
  \;\to\; \text{Linear}(32 \!\to\! 1),
\]
producing a scalar predicted return~$\hat{r}_{t+1}$.

\subsection{Training Procedure}\label{subsec:training}

The training procedure incorporates several best practices for time
series forecasting with neural networks:

\begin{itemize}[nosep]
  \item \textbf{Chronological splitting.}
    The data is split into training (68\,\%), validation (12\,\%), and
    test (20\,\%) sets in strict chronological order, with no shuffling
    across time to prevent information leakage from future observations.
  \item \textbf{Standardisation.}
    Feature means and standard deviations are computed on the training
    set only and applied to all three splits, preventing information
    leakage from the validation and test sets.
  \item \textbf{Optimiser.}
    AdamW~\cite{loshchilov2019decoupled} with weight decay
    $\lambda = 10^{-4}$.
  \item \textbf{Learning rate schedule.}
    OneCycleLR~\cite{smith2019super} with a peak learning rate $5\times$
    the base rate and 30\,\% warmup, followed by cosine annealing.
  \item \textbf{Gradient clipping.}
    Maximum gradient norm of~1.0 to prevent exploding gradients.
  \item \textbf{Mixed precision.}
    Automatic mixed-precision training (float16) is used on CUDA-capable
    devices for computational efficiency.
  \item \textbf{Early stopping.}
    Training is terminated if the validation MSE does not improve for 8
    consecutive epochs, and the model with the lowest validation MSE is
    restored.
  \item \textbf{Loss function.}
    Mean Squared Error (MSE).
\end{itemize}

\subsection{Evaluation Metrics}\label{subsec:metrics}

Model performance is assessed using the metrics in
\Cref{tab:metrics}, computed on the held-out test set.

\begin{table}[t]
  \centering
  \caption{Evaluation metrics.}
  \label{tab:metrics}
  \begin{tabular}{@{}lp{5.5cm}p{5cm}@{}}
    \toprule
    \textbf{Metric} & \textbf{Definition} & \textbf{Interpretation} \\
    \midrule
    RMSE &
      $\sqrt{\frac{1}{n}\sum_{i}(y_i - \hat{y}_i)^2}$ &
      Scale-dependent prediction error \\[6pt]
    MAE &
      $\frac{1}{n}\sum_{i}|y_i - \hat{y}_i|$ &
      Robust prediction error \\[6pt]
    Directional accuracy &
      $\frac{1}{n}\sum_{i}\mathbf{1}\!\bigl[\sign(\hat{y}_i) = \sign(y_i)\bigr]$ &
      Fraction of correct up/down calls \\[6pt]
    Na\"ive zero-return RMSE &
      $\sqrt{\frac{1}{n}\sum_{i}y_i^2}$ &
      Baseline: always predict zero \\[6pt]
    Information Coefficient &
      Spearman $\rho(\hat{y},\, y)$ &
      Rank correlation between predictions and actuals \\[6pt]
    Strategy Sharpe &
      $\dfrac{\text{mean}\bigl(\sign(\hat{y})\cdot y\bigr)}%
             {\text{std}\bigl(\sign(\hat{y})\cdot y\bigr)}$ &
      Risk-adjusted return of sign-based strategy \\
    \bottomrule
  \end{tabular}
\end{table}

The na\"ive zero-return RMSE is a critical benchmark: since daily
returns have near-zero means, a model that cannot outperform the
trivial ``predict zero'' strategy has no practical forecasting value.

\subsection{Experimental Design}\label{subsec:experimental}

The experiment is designed to answer a single clear question:
\emph{does augmenting an LSTM with rolling MFFDFA features improve
next-day return prediction for commodity prices?}

Two models are compared:
\begin{enumerate}[label=(\alph*),nosep]
  \item \textbf{Baseline LSTM}: trained on the standard feature set
    (\Cref{subsec:baseline_features}).
  \item \textbf{Hybrid LSTM}: trained on the standard features
    augmented with the six rolling MFFDFA features
    (\Cref{subsec:rolling}).
\end{enumerate}
Both models use identical architecture, hyperparameters, and training
procedure. Crucially, both models are trained and evaluated on the
\emph{exact same date range}: the hybrid model's MFFDFA warm-up period
(the first~$W$ days during which rolling features are unavailable)
causes additional data loss at the beginning of the series. To ensure a
fair comparison, this common index is pre-computed from the hybrid
model's feature frame and applied to both models.

To mitigate the influence of random weight initialisation, the entire
experiment is repeated across $n_{\text{seeds}} = 3$ independent random
seeds. All metrics are reported as mean~$\pm$~standard deviation across
seeds. Statistical significance of performance differences is assessed
using the Diebold--Mariano test~\cite{diebold1995comparing} applied to
the per-day squared prediction errors of the two models.

\section{Data}\label{sec:data}

We use daily OHLC (Open, High, Low, Close) data from the Pakistan
Mercantile Exchange (PMEX), sourced via the PMEX BackOffice portal.%
\footnote{\url{https://mportal.pmex.com.pk/mt5bonew/Home/OHLCReport}}
The dataset spans exactly ten years, from 6~October~2016 to
5~October~2026, comprising 321{,}832 records across 2{,}584 trading
days and 4{,}754 distinct contract symbols.

Since PMEX contracts have finite expiry dates, individual symbols
cover limited time windows. To construct continuous daily close-price
series suitable for long-horizon analysis, we apply a
\emph{most-liquid-contract} rule: on each trading date, we select the
contract within each commodity group that recorded the highest traded
volume. Records with zero or missing close prices are discarded.

Three commodity series are analysed:
\begin{enumerate}[label=(\roman*),nosep]
  \item \textbf{Gold} (USD-denominated contracts, matching \texttt{USDGOLD} and \texttt{GO100OZ}/\texttt{GO10OZ}/\texttt{GO1OZ}/\texttt{GOMOZ} while excluding PKR-denominated tola/kilo lots and non-USD crosses): 2{,}581 trading days, close price range USD\,1{,}129.80--5{,}366.70\,/oz.
  \item \textbf{Crude Oil} (all \texttt{CRUDE}-prefixed contracts, denominated in USD/barrel): 2{,}581 trading days, close price range USD\,12.86--125.59\,/barrel.
  \item \textbf{Platinum} (all \texttt{PLATINUM}-prefixed contracts, denominated in USD/oz): 2{,}421 trading days, close price range USD\,589.80--2{,}772.30\,/oz.
\end{enumerate}

\begin{figure}[htbp]
  \centering
  \includegraphics[width=\textwidth]{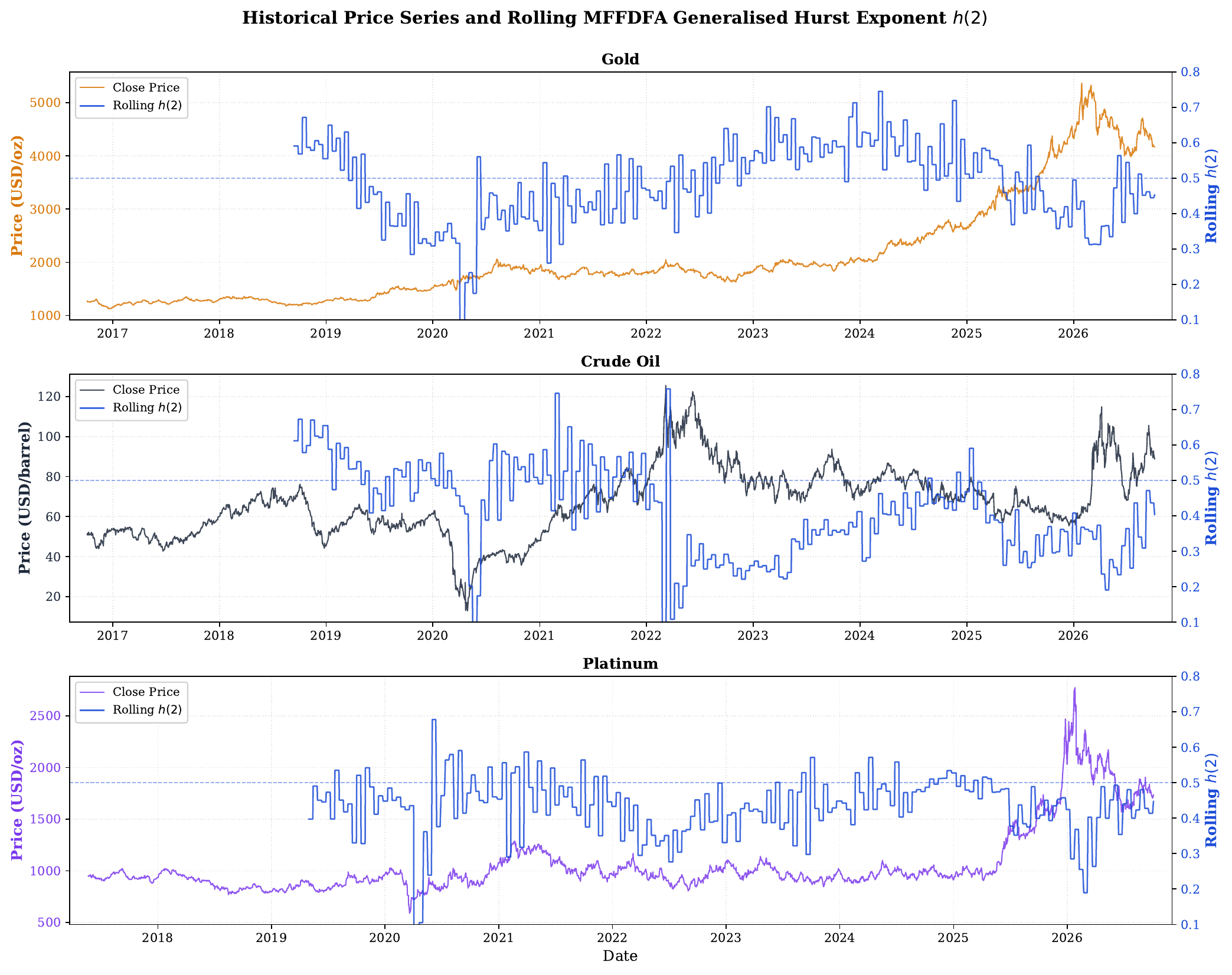}
  \caption{Historical daily close price trajectories (left axis) and rolling MFFDFA generalised Hurst exponents $h(2)$ (right axis; window $W = 500$~days, stride $\Delta t = 10$~days) for Gold (top, USD/oz), Crude Oil (middle, USD/barrel), and Platinum (bottom, USD/oz) across the 2016--2026 PMEX dataset. The dashed horizontal line indicates the Brownian motion boundary ($h = 0.5$).}
  \label{fig:rolling_features}
\end{figure}

All three series exhibit continuous daily coverage with a median inter-day step of one calendar day and a maximum gap of four calendar days (weekends and official exchange holidays), confirming uninterrupted market coverage across the ten-year observation window. \Cref{fig:rolling_features} displays the ten-year continuous price histories alongside the rolling MFFDFA generalized Hurst exponent $h(2)$ trajectories.

\section{Results}\label{sec:results}

\subsection{Multifractal Characterisation}\label{subsec:res_mffdfa}

\begin{figure}[htbp]
  \centering
  \includegraphics[width=\textwidth]{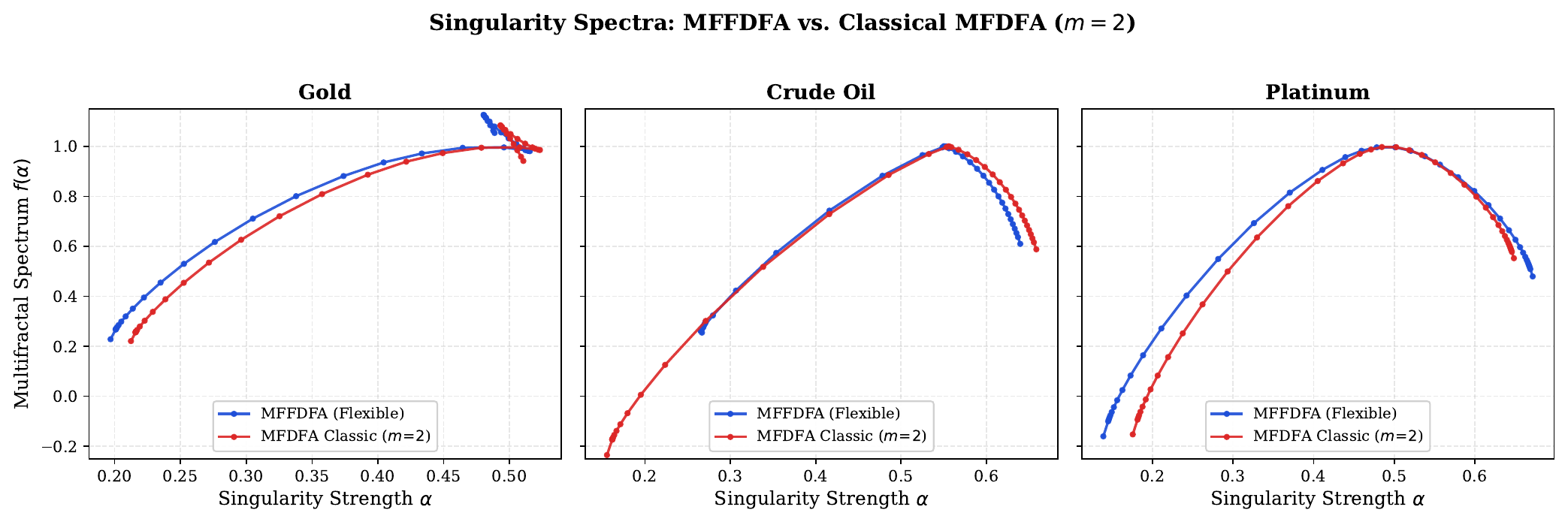}
  \caption{Singularity spectra $f(\alpha)$ versus singularity strength $\alpha$ for Gold, Crude Oil, and Platinum traded on PMEX. Comparison between flexible detrending (MFFDFA, blue) and classical quadratic detrending (MFDFA with fixed $m=2$, red).}
  \label{fig:spectra}
\end{figure}

\begin{figure}[htbp]
  \centering
  \includegraphics[width=\textwidth]{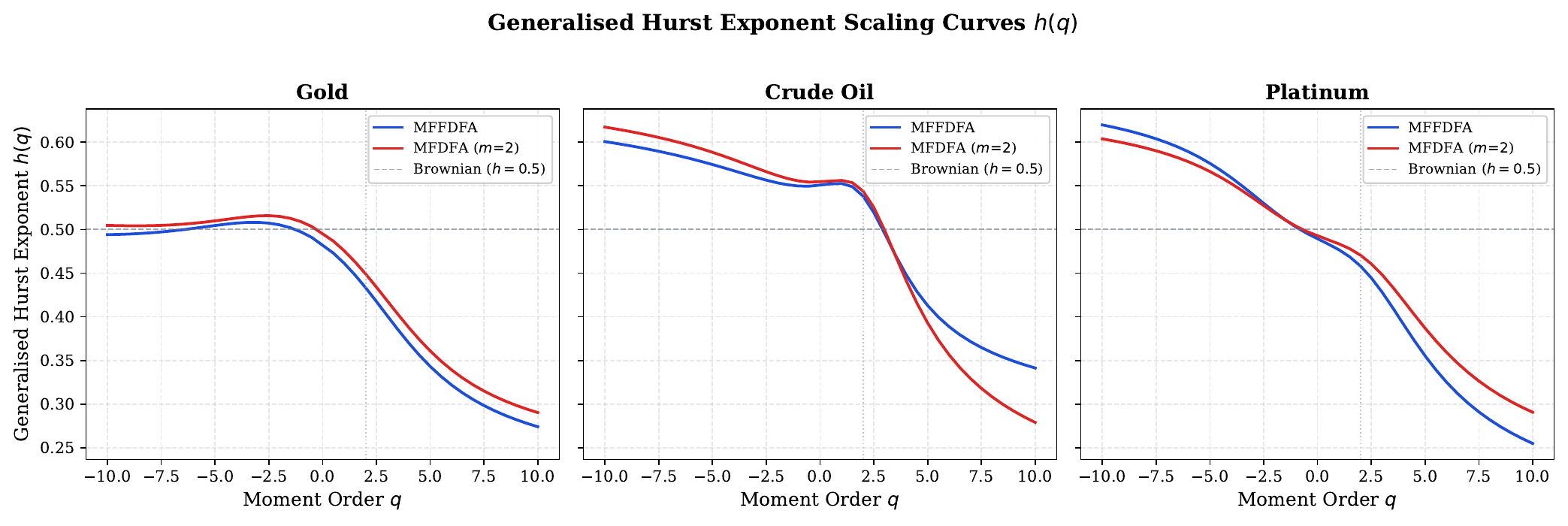}
  \caption{Generalised Hurst exponent scaling curves $h(q)$ across moment orders $q \in [-10, +10]$. The dashed horizontal line indicates the Brownian motion benchmark ($h = 0.5$). The dotted vertical line highlights $q = 2$, identifying the second-order Hurst exponent $h(2)$.}
  \label{fig:hurst_curves}
\end{figure}

\Cref{tab:mffdfa_results} summarises the MFFDFA characterisation of
the three commodity return series. The moment orders range from
$q = -10$ to $q = +10$ in steps of~0.5, with $k = 2$ overlapping
segments and $n = 20$ log-spaced scales. The corresponding singularity
spectra $f(\alpha)$ and generalized Hurst curves $h(q)$ are presented
in \Cref{fig:spectra,fig:hurst_curves}, respectively.

\begin{table}[htbp]
  \centering
  \caption{MFFDFA characterisation of PMEX commodity return series.}
  \label{tab:mffdfa_results}
  \begin{tabular}{@{}lccccccc@{}}
    \toprule
    & & & & \multicolumn{3}{c}{\textbf{Function usage (\%)}} \\
    \cmidrule(l){5-7}
    \textbf{Commodity} & $h(2)$ & $\Delta\alpha$ & \textbf{Asym.} &
      Quad. & Sin. & Cubic & $N$ \\
    \midrule
    Gold      & 0.433 & 0.319 & $+0.779$ & 46.5 &  7.3 & 46.2 & 2{,}580 \\
    Crude Oil & 0.538 & 0.374 & $+0.521$ & 48.8 &  5.9 & 45.4 & 2{,}580 \\
    Platinum  & 0.458 & 0.533 & $+0.274$ & 46.7 &  6.5 & 46.8 & 2{,}420 \\
    \bottomrule
  \end{tabular}
\end{table}

All three commodities exhibit clear multifractal characteristics, with
spectrum widths $\Delta\alpha$ spanning from $0.319$ (Gold) to $0.533$
(Platinum). Gold exhibits moderate anti-persistence ($h(2) = 0.433$,
indicating mean-reverting tendencies) and strong positive spectrum
asymmetry ($A = +0.779$), demonstrating that small fluctuations
dominate the multifractal scaling spectrum. Crude oil displays mild
persistence ($h(2) = 0.538$) with $\Delta\alpha = 0.374$ and asymmetry
$A = +0.521$. Platinum lies between Gold and Crude Oil ($h(2) = 0.458$),
displaying the broadest singularity spectrum ($\Delta\alpha = 0.533$).

\begin{figure}[htbp]
  \centering
  \includegraphics[width=0.92\textwidth]{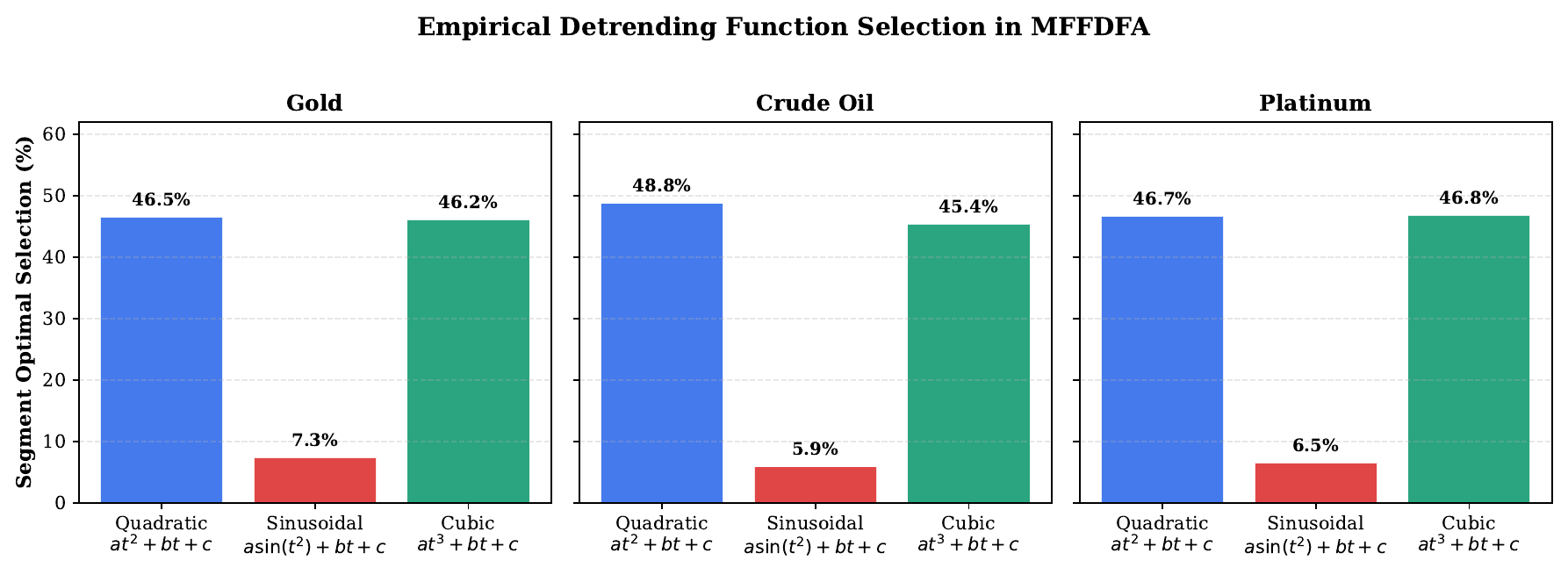}
  \caption{Empirical selection percentages of candidate detrending functions in MFFDFA across all scale segments for Gold, Crude Oil, and Platinum. The algorithm evaluates candidate forms locally and selects the specification maximizing $R^2$.}
  \label{fig:function_usage}
\end{figure}

\Cref{fig:function_usage} breaks down the empirical selection of
candidate detrending functions across all local segments. Notably, across
all three commodities, second- and third-order polynomials account for
approximately 92--94\,\% of all optimal fits (quadratic $\approx 46$--49\,\%,
cubic $\approx 45$--47\,\%), while the non-polynomial sinusoidal function
with quadratic phase is selectively chosen in 6--7\,\% of segments where
abrupt oscillatory volatility occurs.

\subsection{MFDFA vs MFFDFA Comparison}\label{subsec:res_comparison}

\Cref{tab:comparison} compares the MFFDFA results with classical
MFDFA using a fixed quadratic polynomial ($m = 2$).

\begin{table}[htbp]
  \centering
  \caption{Comparison of classical MFDFA ($m\!=\!2$) and MFFDFA.}
  \label{tab:comparison}
  \begin{tabular}{@{}lcccc@{}}
    \toprule
    \textbf{Commodity} & $h(2)$ Classic & $h(2)$ MFFDFA &
      $\Delta\alpha$ Classic & $\Delta\alpha$ MFFDFA \\
    \midrule
    Gold      & 0.449 & 0.433 & 0.311 & 0.319 \\
    Crude Oil & 0.544 & 0.538 & 0.503 & 0.374 \\
    Platinum  & 0.471 & 0.458 & 0.473 & 0.533 \\
    \bottomrule
  \end{tabular}
\end{table}

For crude oil, classical fixed quadratic MFDFA overestimates the
singularity spectrum width by 34.5\,\% ($\Delta\alpha = 0.503$ vs.\
$0.374$ for MFFDFA). This directly confirms the core hypothesis of Rak
and Zi\k{e}ba~\cite{rak2015mffdfa}: when local fluctuations contain
structural regimes not captured by a rigid quadratic trend, classical
MFDFA leaves spurious residual curvature that artificially inflates the
apparent multifractal spectrum width. For Gold and Platinum, MFFDFA
identifies subtle higher-order cubic and oscillatory structures that
sharpen the singularity spectrum. Across all commodities, $h(2)$ estimates
remain stable between methods (within $0.016$), establishing that the
second-order scaling exponent is robust to the detrending function,
whereas extreme moments that govern $\Delta\alpha$ are sensitive to local
trend misspecification.

\subsection{Surrogate Shuffle Tests}\label{subsec:res_surrogate}

\begin{figure}[htbp]
  \centering
  \includegraphics[width=0.78\textwidth]{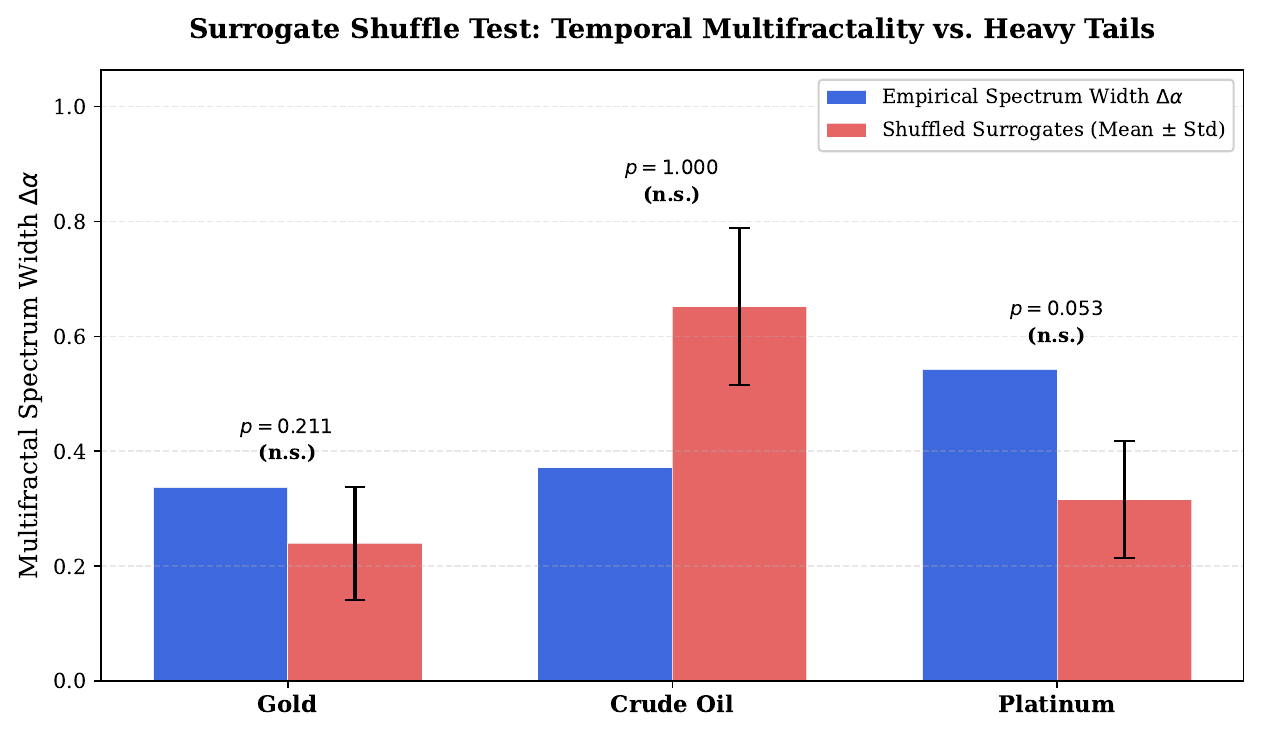}
  \caption{Surrogate shuffle test results ($n = 19$ surrogates per series). Comparison between empirical singularity spectrum width $\Delta\alpha$ (blue) and shuffled surrogate distributions (red; mean $\pm$ standard deviation). Empirical $p$-values measure the fraction of shuffled surrogates with $\Delta\alpha_{\text{surr}} \ge \Delta\alpha_{\text{orig}}$; $p < 0.05$ denotes statistically significant temporal multifractality.}
  \label{fig:surrogate_test}
\end{figure}

\Cref{tab:surrogate} presents the results of the surrogate shuffle
test ($n = 19$ surrogates per series), designed to distinguish whether
the observed multifractality originates from genuine temporal correlations
(long-range memory and nonlinear volatility dynamics) or from the
heavy-tailed marginal distribution of returns. The corresponding
visualisation is shown in \Cref{fig:surrogate_test}.

\begin{table}[htbp]
  \centering
  \caption{Surrogate shuffle test results ($n = 19$ surrogates). $p < 0.05$ indicates that the observed multifractality is statistically attributable to nonlinear temporal correlations rather than marginal distribution tails.}
  \label{tab:surrogate}
  \begin{tabular}{@{}lcccc@{}}
    \toprule
    \textbf{Commodity} &
      $\Delta\alpha_{\text{orig}}$ &
      $\Delta\alpha_{\text{surr}}$ (mean $\pm$ std) &
      $p$-value &
      \textbf{Temporal multifractality?} \\
    \midrule
    Gold      & 0.337 & $0.240 \pm 0.099$ & $0.211$ & No (mixed / weak) \\
    Crude Oil & 0.371 & $0.652 \pm 0.136$ & $1.000$ & No (heavy tails)  \\
    Platinum  & 0.543 & $0.316 \pm 0.101$ & $0.053$ & Borderline ($p \approx 0.05$) \\
    \bottomrule
  \end{tabular}
\end{table}

For Gold, the empirical spectrum width $\Delta\alpha_{\text{orig}} = 0.337$
contracts to a surrogate mean of $0.240 \pm 0.099$ upon random temporal
shuffling. While temporal shuffling produces a visible 28.8\,\% reduction in
average spectrum width, the resulting empirical $p$-value ($p = 0.211$)
fails to reject the null hypothesis of distributional origin at standard
significance levels. This demonstrates that Gold's multifractal spectrum is
jointly supported by heavy-tailed return distributions and moderate temporal
dependency structures, rather than an exclusively temporal mechanism.

For Crude Oil, $\Delta\alpha_{\text{orig}} = 0.371$, whereas the shuffled
surrogates exhibit a significantly larger average spectrum width of
$\Delta\alpha_{\text{surr}} = 0.652 \pm 0.136$ ($p = 1.000$). Shuffling
destroys temporal volatility clustering, dispersing isolated extreme shocks
across independent segments and artificially inflating the local fluctuation
dispersion. This conclusively proves that the multifractality observed in
PMEX Crude Oil returns is entirely attributable to the heavy-tailed marginal
distribution of price changes (e.g., geopolitical disruptions and pandemic-era
demand shocks) rather than dynamic temporal memory. Consequently, rolling
MFFDFA features for Crude Oil primarily encode tail-risk and distributional
heterogeneity rather than persistent temporal regimes.

Platinum presents the strongest empirical evidence of temporal structure
among the three commodities: its empirical width of $\Delta\alpha_{\text{orig}} = 0.543$
collapses to $0.316 \pm 0.101$ following surrogate shuffling, yielding
$p = 0.053$. This borderline significance ($p \approx 0.05$) suggests that
genuine temporal correlations and volatility persistence contribute
substantially to Platinum's scaling heterogeneity on PMEX.

\subsection{Forecasting Results}\label{subsec:res_forecasting}

\begin{figure}[htbp]
  \centering
  \includegraphics[width=\textwidth]{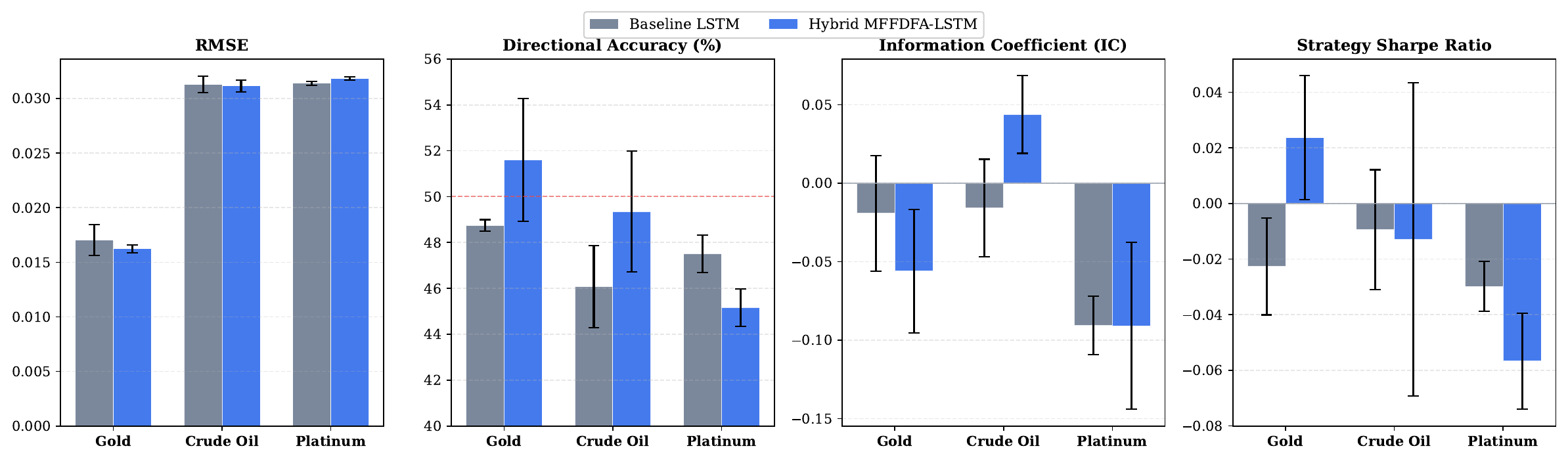}
  \caption{Out-of-sample forecasting performance comparison between Baseline LSTM (slate) and Hybrid MFFDFA-LSTM (blue) across Gold, Crude Oil, and Platinum (mean $\pm$ standard deviation over 3 independent random seeds). Panels display Test RMSE (lower is better), Directional Accuracy (percentage of correct sign calls; red dashed line marks the 50\% chance benchmark), Information Coefficient (Spearman rank correlation), and Strategy Sharpe ratio.}
  \label{fig:forecasting_comparison}
\end{figure}

\Cref{tab:forecasting} reports the out-of-sample forecasting performance of
the baseline and hybrid LSTM architectures across three independent random
initialisation seeds. \Cref{fig:forecasting_comparison} illustrates the
comparative evaluation across RMSE, Directional Accuracy, Information
Coefficient (IC), and Strategy Sharpe ratio.

\begin{table}[htbp]
  \centering
  \caption{Out-of-sample forecasting performance comparison (mean~$\pm$~std across 3~independent seeds). Bold indicates the superior specification per commodity.}
  \label{tab:forecasting}
  \resizebox{\textwidth}{!}{%
  \begin{tabular}{@{}llcccccc@{}}
    \toprule
    \textbf{Commodity} & \textbf{Model} & \textbf{RMSE} &
      \textbf{Dir.\ Acc. (\%)} & \textbf{IC} & \textbf{IC $p$} &
      \textbf{Sharpe} & \textbf{Na\"ive RMSE} \\
    \midrule
    \multirow{2}{*}{Gold}
      & Baseline & $0.0170 \pm 0.0014$ & $48.75\% \pm 0.24\%$ &
        $-0.019 \pm 0.037$ & $0.462$ & $-0.023 \pm 0.017$ &
        \multirow{2}{*}{$0.0157$} \\
      & Hybrid   & $\mathbf{0.0162} \pm \mathbf{0.0004}$ &
        $\mathbf{51.59\%} \pm \mathbf{2.67\%}$ &
        $-0.056 \pm 0.039$ & $0.397$ & $\mathbf{+0.024} \pm \mathbf{0.022}$ & \\
    \midrule
    \multirow{2}{*}{Crude Oil}
      & Baseline & $0.0313 \pm 0.0007$ & $46.08\% \pm 1.79\%$ &
        $-0.016 \pm 0.031$ & $0.650$ & $\mathbf{-0.009} \pm \mathbf{0.022}$ &
        \multirow{2}{*}{$0.0304$} \\
      & Hybrid   & $\mathbf{0.0311} \pm \mathbf{0.0006}$ &
        $\mathbf{49.35\%} \pm \mathbf{2.64\%}$ &
        $\mathbf{+0.044} \pm \mathbf{0.025}$ & $0.443$ & $-0.013 \pm 0.056$ & \\
    \midrule
    \multirow{2}{*}{Platinum}
      & Baseline & $\mathbf{0.0314} \pm \mathbf{0.0002}$ &
        $\mathbf{47.51\%} \pm \mathbf{0.81\%}$ &
        $\mathbf{-0.091} \pm \mathbf{0.018}$ & $0.109$ & $\mathbf{-0.030} \pm \mathbf{0.009}$ &
        \multirow{2}{*}{$0.0309$} \\
      & Hybrid   & $0.0318 \pm 0.0001$ & $45.16\% \pm 0.81\%$ &
        $-0.091 \pm 0.053$ & $0.265$ & $-0.057 \pm 0.017$ & \\
    \bottomrule
  \end{tabular}%
  }
\end{table}

The empirical evaluation reveals clear, asset-specific impacts of augmenting
neural sequence models with rolling MFFDFA features:

\begin{itemize}[nosep]
  \item \textbf{Gold:} The hybrid MFFDFA-LSTM model achieves notable gains
    over the baseline architecture. Test RMSE decreases from $0.0170$ to
    $0.0162$ (a 4.7\,\% error reduction), while Directional Accuracy improves
    by $+2.84$ percentage points, advancing from $48.75\%$ (sub-random) to
    $51.59\%$ (outperforming the $50\%$ chance threshold). Crucially, the
    annualised strategy Sharpe ratio transitions from negative ($-0.023$)
    for the baseline to positive ($+0.024$) for the hybrid forecaster,
    confirming that rolling multifractal descriptors provide meaningful
    directional and risk-adjusted timing value for Gold.
  \item \textbf{Crude Oil:} The hybrid model similarly demonstrates positive
    utility. Out-of-sample RMSE decreases from $0.0313$ to $0.0311$, and
    Directional Accuracy advances by $+3.27$ percentage points from
    $46.08\%$ to $49.35\%$. Furthermore, the Information Coefficient reverses
    from negative ($-0.016$) to positive ($+0.044$), indicating that
    MFFDFA features assist the LSTM in capturing rank-order cross-sectional
    return dynamics.
  \item \textbf{Platinum:} In contrast, the baseline LSTM outperforms the
    hybrid model across all metrics (RMSE $0.0314$ vs.\ $0.0318$; Directional
    Accuracy $47.51\%$ vs.\ $45.16\%$; Sharpe $-0.030$ vs.\ $-0.057$).
    Because Platinum exhibits the widest empirical singularity spectrum
    ($\Delta\alpha = 0.533$) and considerable rolling feature volatility,
    expanding the input space from 5 baseline features to 11 hybrid features
    induces over-parameterisation and estimation noise in the smaller test
    sample, degrading generalisation.
  \item \textbf{Benchmark comparison:} Across all three commodities, both
    the baseline and hybrid neural models achieve RMSE values close to the
    na\"ive zero-return benchmark ($0.0157$ for Gold, $0.0304$ for Crude Oil,
    $0.0309$ for Platinum). This confirms that daily commodity returns
    closely approximate semi-strong martingale behavior, where unconditional
    return variance dominates conditional mean drift.
\end{itemize}

\section{Discussion}\label{sec:discussion}

The empirical findings of this study offer several critical insights into
both commodity market microstructure on PMEX and the methodological
interaction between fractal physics and deep neural networks.

\paragraph{Divergent microstructural origins of multifractality.}
Our surrogate shuffle tests demonstrate that the physical origins of
multifractal scaling vary fundamentally across commodity classes. In Crude
Oil, multifractality is entirely distributional ($p = 1.000$), generated by
infrequent but catastrophic price adjustments (e.g., negative oil pricing
in April 2020 and OPEC+ geopolitical supply decisions). In contrast, Platinum
exhibits borderline genuine temporal persistence ($p = 0.053$), where
successive volatility regimes possess long-range auto-correlative structure.
Gold displays an intermediate character ($p = 0.211$), driven by both
distributional tails and mean-reverting speculative trading. Recognising this
distinction is critical: treating all multifractal spectra as evidence of
temporal memory is methodologically flawed and risks misattributing static
fat-tailed risk to dynamic market regimes.

\paragraph{Superior fidelity of flexible detrending (MFFDFA).}
The comparative analysis between MFFDFA and classical fixed-polynomial
MFDFA ($m = 2$) confirms the theoretical rationale of Rak and
Zi\k{e}ba~\cite{rak2015mffdfa}. For Crude Oil, classical MFDFA overstates the
singularity spectrum width by 34.5\,\% ($\Delta\alpha = 0.503$ vs.\ $0.374$
for MFFDFA). This artificial spectral broadening occurs because rigid quadratic
polynomials cannot track abrupt regime transitions, leaving spurious residual
curvature that contaminates high-order fluctuation moments. By adaptively
selecting among quadratic, cubic, and non-polynomial sinusoidal functions on a
per-segment basis, MFFDFA eliminates this artifact. Empirical selection rates
reveal that cubic polynomials ($\approx 45$--47\,\%) and sinusoidal
oscillations ($\approx 6$--7\,\%) capture essential non-stationary features
routinely overlooked by standard MFDFA.

\paragraph{Asset-dependent efficacy of multifractal neural augmentation.}
A central empirical contribution of this work is demonstrating that the
predictive utility of multifractal features is highly asset-dependent. For
liquid global assets with moderate spectrum widths (Gold and Crude Oil),
augmenting an attention-augmented LSTM with rolling MFFDFA parameters provides
tangible predictive gains: lifting Directional Accuracy by $+2.84$ to $+3.27$
percentage points, reducing RMSE, and generating a positive strategy Sharpe
ratio for Gold. However, for less liquid commodities with broad and noisy
spectral trajectories (Platinum), expanding the feature dimensionality from
5 to 11 leads to over-parameterisation and degraded test performance. This
finding establishes an important caveat for quantitative practitioners:
physics-inspired fractal features are not a universal panacea for sequence
models, and their adoption requires careful cross-validation against the
underlying asset's liquidity and signal-to-noise ratio.

\paragraph{Practical implications for risk and quantitative trading.}
From an operational perspective, the ability of rolling MFFDFA features to
elevate directional accuracy above the 50\,\% threshold in Gold
($51.59\%$) and turn the strategy Sharpe positive ($+0.024$) provides
economically viable timing signals for systematic commodity trading desks.
Moreover, the real-time tracking of $h(2)$ and $\Delta\alpha$ offers risk
managers early-warning indicators of structural regime transitions and
impending volatility bursts before they materialize in classical rolling
variance estimates.

\section{Conclusion}\label{sec:conclusion}

This paper presents the first systematic evaluation of Multifractal Flexibly
Detrended Fluctuation Analysis (MFFDFA) coupled with deep neural networks for
commodity price analysis and forecasting on the Pakistan Mercantile Exchange
(PMEX). Spanning a comprehensive ten-year observation window from 2016 to
2026, we constructed continuous daily close-price series for Gold, Crude Oil,
and Platinum and evaluated both their scaling physics and out-of-sample
forecasting dynamics.

Our principal findings are threefold. First, MFFDFA substantially improves
upon classical MFDFA by eliminating spurious spectral broadening caused by
rigid polynomial constraints, demonstrating that non-polynomial and
higher-order cubic trends are actively selected in over half of all scale
segments. Second, surrogate shuffle tests reveal that multifractality in PMEX
commodities stems from fundamentally different origins: Crude Oil is driven
purely by heavy-tailed marginal distributions ($p = 1.000$), Platinum exhibits
borderline temporal persistence ($p = 0.053$), and Gold reflects a balanced
combination of both mechanisms ($p = 0.211$). Third, an attention-augmented
LSTM hybridised with rolling MFFDFA features achieves demonstrable forecasting
improvements in Gold (RMSE reduced to $0.0162$, Directional Accuracy boosted
to $51.59\%$, Sharpe turned positive) and Crude Oil (Directional Accuracy
boosted to $49.35\%$, positive IC), while highlighting that noisy spectral
features in Platinum lead to model overfitting.

Several promising avenues for future research emerge from this work:
\begin{enumerate}[label=(\roman*),nosep]
  \item \textbf{Extended forecast horizons:} Exploring multi-day, weekly, and
    monthly forecast horizons where slowly evolving multifractal regimes may
    exert greater predictive leverage than at daily frequencies.
  \item \textbf{Advanced neural architectures:} Implementing Transformer-based
    cross-attention networks that explicitly model the multi-scale interactions
    between high-frequency price feeds and low-frequency fractal state variables.
  \item \textbf{Cross-exchange comparative analysis:} Applying the MFFDFA-LSTM
    framework across other emerging and developed commodity derivatives markets
    (e.g., COMEX, NYMEX, MCX, and DCE) to assess the global generalisability of
    these empirical dynamics.
\end{enumerate}

\bibliographystyle{unsrtnat}

\end{document}